\documentclass[useAMS,usenatbib]{mn2e}
\pdfoutput=1
\usepackage{graphicx}
\usepackage{dcolumn}
\usepackage{bm}
\usepackage{amssymb,amsmath}
\usepackage{color}
\usepackage{fixltx2e}
\usepackage{hyperref}
\hypersetup{colorlinks=true,citecolor=blue}
\usepackage{epsfig}
\usepackage{placeins}
\usepackage{array}
\usepackage{natbib}
\usepackage{mathrsfs}
\usepackage{multirow}
\usepackage{booktabs}
\bibpunct{(}{)}{;}{a}{}{,}
\newcommand{\nv}{\hat{\bf n}}
\newcommand{\be}{\begin{equation}}
\newcommand{\ee}{\end{equation}}
\newcommand{\ba}{\begin{eqnarray}}
\newcommand{\ea}{\end{eqnarray}}
\newcommand{\bal}{\begin{align}}
\newcommand{\eal}{\end{align}}
\newcommand{\bfig}{\begin{figure}}
\newcommand{\efig}{\end{figure}}

\newcommand{\planck}{{\sl{Planck}}}
\newcommand{\wmap}{{\sl{WMAP}}}
\newcommand  \gtsim  {\lower.5ex\hbox{$\; \buildrel > \over \sim \;$}}
\newcommand  \ltsim  {\lower.5ex\hbox{$\; \buildrel < \over \sim \;$}}
\newcommand{\vecn}{\hat{\mathbf{n}}}

\title[The Python Sky Model: software for simulating the Galactic microwave sky]
{The Python Sky Model: software for simulating the Galactic microwave sky}
\author[B.~Thorne, J.~Dunkley, D.~Alonso]
{B.~Thorne$^{1}$, J.~Dunkley$^{1}$, D.~Alonso$^{1}$, S.~N\ae ss$^{1}$ \\
 $^1$Department of Physics, University of Oxford, Keble Road, Oxford OX1 3RH
}

\begin{document}
 \date{\today}
 \pagerange{1--12} \pubyear{2016}
 \maketitle
 
\begin{abstract}
We present a numerical code to simulate maps of Galactic emission in intensity and polarization at 
microwave frequencies, aiding in the design of Cosmic Microwave Background experiments.  This Python code 
builds on existing efforts to simulate the sky by providing an easy-to-use interface and is based 
on publicly available data from the \wmap\ and \planck\ satellite missions. We simulate 
synchrotron, thermal dust, free-free, and anomalous microwave emission over the whole sky, in 
addition to the Cosmic Microwave Background, and include a set of alternative prescriptions for the 
frequency dependence of each component that are consistent with current data. We also present a 
prescription for adding small-scale realizations of these components at resolutions 
greater than current all-sky measurements. The code is available at \url{https://github.com/bthorne93/PySM_public}.
\end{abstract}

\section{Introduction}
\label{sec:intro}

In recent years the temperature and polarization anisotropies of the Cosmic Microwave Backround 
(CMB) have been measured with increasing precision by the \wmap\ and \planck\ satellites 
\citep{Hinshaw2013,planckmission:2015}, coupled with ground and balloon-based observations. The 
constraints these observations place on the parameters that describe the cosmology of the Universe 
have been tight enough to usher in the era of `precision cosmology'.  

Further progress will be made by measuring the polarization anisotropies of the CMB to greater 
precision.  It is in the power spectrum of these anisotropies that the signature of primordial 
gravitational waves may be found, which would provide strong evidence in support of the scenario 
that the Universe went through an early period of inflation \citep[e.g.,][]{Baumann2009}. Current
polarization data are starting to provide the strongest constraints on primordial gravitational
waves \citep{2015PhRvL.114j1301B}.

The CMB temperature anisotropy dominates over foreground emission from the Galaxy in a broad range 
of frequencies.  In contrast, the polarized CMB signal is weaker than the strongly polarized 
Galactic thermal dust and synchrotron radiation. In particular, the divergence-free $B$-mode 
polarization signal sourced by primordial gravitational waves at recombination is predicted to be 
at least several orders of magnitude weaker than the polarized foregrounds, averaged over the sky, 
and is a subdominant signal even in the cleanest sky regions \citep{planck_dust:2015}.  

To optimize our ability to extract the CMB polarization signal from upcoming and future experiments 
 we rely on realistic models of the Galactic emission to simulate observations of these components 
at a range of frequencies.  Several sky simulation tools are already publically available, 
including the Planck Sky Model \citep{Delabrouille2012} and the Global Sky Model
\citep{DeOliveira-Costa2008,Zheng2016}. While our work was in 
preparation a similar modeling and software effort was presented in \citet{HerviasCaimapo2016} for 
polarized Galactic emission.  

With the new code presented here we build on existing efforts, providing a flexible and easily used 
tool for simulating Galactic emission that includes recent public data from the \planck\ satellite. 
We do not attempt to physically model the emission in three dimensions, via for example, 
integrating a dust or electron density over a Galactic magnetic field 
\citep[e.g.][]{waelkens/etal:2009,fauvet/etal:2011,fauvet/etal:2012,jansson/farrar:2012,orlando/strong:2013,beck/etal:2014}. 
Instead we adopt empirical models that describe the frequency scaling of 
each component with simple forms consistent with current data, using high signal-to-noise maps of 
each component as templates at frequencies far from the foreground minimum. These simulations are 
therefore limited in scope and will not capture the complexity present in the true emission. 

The structure of this paper is as follows:  in \S\ref{sec:lss} we describe the structure of the 
code together with the models and alternatives used for each component. In \S\ref{sec:smallscale} 
we describe a procedure to add small-scale anisotropy to the simulated maps; and in 
\S\ref{sec:discussion} we summarize the usefulness and limitations of these simulations.  

\section{Large-scale simulations}
\label{sec:lss}

We simulate Galactic diffuse emission in intensity and polarization from four Galactic 
components: thermal dust, synchrotron, free-free, and anomalous microwave emission (AME). We also 
include a gravitationally lensed CMB realization and white instrument noise. Maps can be integrated 
over a top-hat bandpass describing the response of each experimental channel, and smoothed with a
Gaussian beam. 

\begin{figure*}
\includegraphics[width=0.96\textwidth]{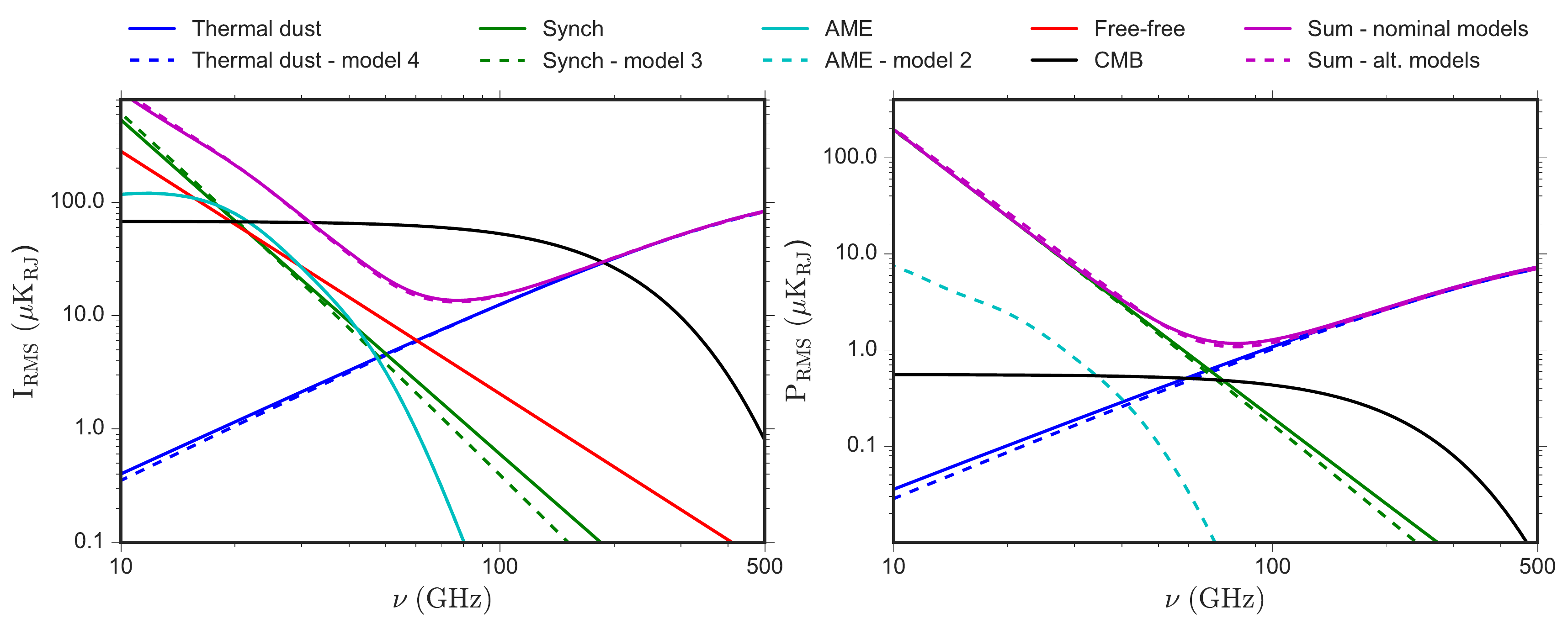}
\caption{The frequency scaling laws for the individual components of {\tt PySM}; we show the nominal and 
alternative models as solid and dashed lines respectively.  We show only the alternative models which have 
a significant impact on the shape of the spectrum.  These spectra are calculated by producing masked 
maps of each component at each frequency, smoothing to FWHM 1$^{\circ}$ in intensity and 40$'$ in 
polarization, and then computing the RMS.  The mask used in intensity is the \wmap\ 9 year KQ85 
mask, and the polarization mask is the \planck\ polarization confidence mask CPM83.}
\label{fig:scalinglaws}
\end{figure*}

\begin{figure*}
\includegraphics[width=\textwidth]{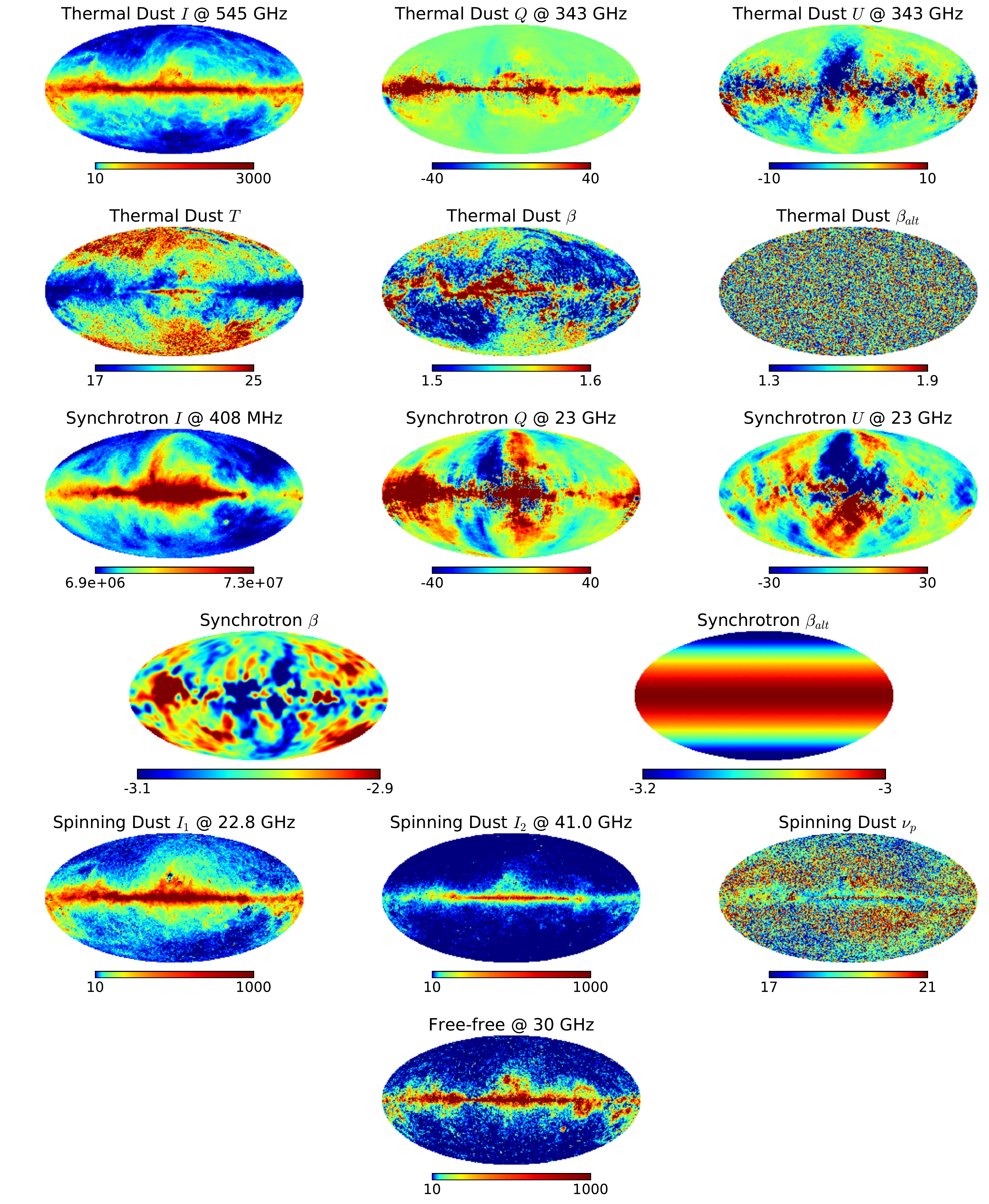}
\caption{Template maps used in the {\tt PySM} models.  All emission templates are in units of $\mu 
\text{K}_{\text{RJ}}$ and all dust temperature templates are in K.  Intensity templates are plotted on a 
log scale, and the polarization templates on a linear scale.}
\label{fig:allmaps}
\end{figure*}

The user specifies a set of observation frequencies, beam widths, bandpass widths, noise and
chosen output components and units.  The code simulates each component at each frequency
using a phenomenological model. One or more emission template maps are defined at pivot frequencies,
and then the extrapolation in frequency is performed using scaling laws and maps of spectral parameters. 
A  lensed CMB realization can be included by calling the {\tt Taylens} software \citep{naess/louis:2013} 
directly, or using a pre-calculated realization.  The software is designed to be 
easily extendable to alternative models or scalings. The intensity and polarized emission as a function 
of frequency for the models we consider is summarized in Figure \ref{fig:scalinglaws}, together with the
template maps in Figure \ref{fig:allmaps}.

\subsection{Synchrotron}
\label{subsec:synch}

Synchrotron radiation is the dominant radiation mechanism in polarization at frequencies $\ltsim 
50$ GHz \citep[e.g.,][]{kogut/etal:2007}. It is produced by cosmic rays spiralling around Galactic 
magnetic fields and radiating.  The power and spectral energy distribution depends on both the 
strength of the local magnetic field, and the energy distribution of the injected cosmic rays. The 
polarization of the radiation depends on the orientation of the intervening magnetic field. The 
predicted dependence of the spectrum on the magnetic field for a population of cosmic rays with 
energy distribution $N(E)\propto E^{-p}$ is, in antenna temperature units:
\be
I_{\nu}\propto B^{\frac{p+1}{2}}\nu^{\beta},
\ee
where $\beta=-\frac{(p+3)}{2}$ \citep{ryb}. The spectral index, $\beta$, is expected to have some 
spatial variability and to vary with frequency.  As synchrotron sources age their spectral energy 
distribution (SED) steepens, since high frequency radiation corresponds to higher energy particles 
which radiate energy away most rapidly.  Along a line of sight there will likely also be multiple 
synchrotron components, and the stacking of their spectra can lead to flattening of the SED.  The 
spectrum can also be flattened through effects of synchrotron self-absorption, which tends to be 
more significant towards the Galactic center.

\subsubsection{Model 1: Nominal index}

The nominal {\tt PySM} model assumes that the synchrotron intensity is a scaling of the 
degree-scale-smoothed 408 MHz Haslam map \citep{haslam/etal:1981,haslam/etal:1982}, reprocessed by 
\citet{Remazeilles2014a}. It models the polarization as a scaling of the \wmap\ 9-year 23 GHz Q and 
U maps \citep{bennett/etal:2013}, smoothed to three degrees. Both of these maps have small scales 
added using the prescription described in \S\ref{sec:smallscale}.

In the nominal model we simulate the spectral index as being a power-law in every direction, such 
that
\be
I^{\text{Synch}}_{\nu}(\vecn) = A_{\nu_{0}}(\vecn)\left(\frac{\nu}{\nu_{0}}\right)^{\beta_{s}(\vecn)}.
\label{eqn:synchmodel}
\ee
As in the nominal Planck Sky Model v1.7.8 simulations, we use the spectral index map from `Model 4' 
of \citet{Miville-Deschenes2008}, calculated from a combination of Haslam and \wmap\ 23 GHz 
polarization data using a model of the Galactic magnetic field. We assume that the index is the 
same in temperature and polarization, although the true sky will most likely be more complicated than this. The template 
maps and index map are shown in Fig \ref{fig:allmaps}.

\subsubsection{Model 2: Spatially steepening index}

The cosmic rays responsible for synchrotron radiation are thought to be energized by processes 
such as supernovae, which are more common in the Galactic plane.  Synchrotron emission 
observed at higher latitudes will therefore likely be produced by older cosmic rays which have 
diffused out of the Galactic plane, and therefore lost more energy. This is expected to result in 
the steepening of the synchrotron spectral index  away from the plane 
\citep{Kogut2012,Ichiki2014}. Evidence for steepening in the polarization emission has been 
seen in \citet{kogut/etal:2007,fuskeland/etal:2014,ruud/etal:2015} using \wmap\ and QUIET data. 

We parameterize the steepening with a smoothly varying index described by a gradient $\delta_\beta$  
that scales with Galactic latitude, $b$, such that $\beta_s = \beta_{s,\rm b=0} + \delta_\beta \sin|b|$. 
In Model 2 we use $\delta_\beta=-0.3$, consistent with \wmap\ polarization data \cite{kogut/etal:2007}.  
The simulated index varies from $\beta_{s}=-3.0$ at the equator to $\beta_{s}=-3.3$ at the poles in both 
intensity and polarization, as shown in Figure \ref{fig:allmaps}.  

\subsubsection{Model 3: Curvature of index}

The synchrotron emission may be better modelled by a curved spectrum that either flattens or 
steepens with frequency. 
Model 3 simulates the steepening or flattening of the spectral index above a frequency, $\nu_{c}$ as:

\be
I^{\text{Synch}}_{\nu}(\vecn) = A_{\nu_{0}}(\vecn)\left(\frac{\nu}{\nu_0}\right)^{\beta_{s}(\vecn)+C\ln(\frac{\nu}{\nu_{c}})}     ,
\label{eq:curvepowerlaw}
\ee
where positive $C$ corresponds to flattening and negative $C$ to steepening. 

\citet{Kogut2012} fits this model to a small patch of sky with ten overlapping radio frequency sky surveys
and \wmap\ 23~GHz data, finding best-fit values of $\beta=-2.64 \pm 0.03, C = -0.052 \pm 0.005$ at 0.31 GHz. 
This corresponds to a steepening of about 0.57 between 408~MHz and 94~GHz. Evaluating the spectral index
at 23~GHz \citet{Kogut2012} finds $\beta_{23} = -3.09 \pm 0.05$.  This is consistent with the index map 
in model 1 which has a mean and standard deviation of $-3.00 \pm 0.06$. Therefore, for simplicity we use 
the same map as model 1 from \citet{Miville-Deschenes2008} for $\beta(\vecn)$, and a baseline curvature 
value of $C = -0.052$ at $\nu_c=23$~GHz. 

\subsection{Thermal dust emission}
\label{subsec:dust}

At frequencies greater than $\approx 70$ GHz the polarized foreground spectrum is dominated by thermal dust 
emission. The dust grains are thought to be a combination of carbonaceous and silicate grains, and 
polycyclic aromatic hydrocarbons (PAHs).  The total emission results from the interaction of these 
species with the interstellar radiation field: the grains are heated by absorption in the optical 
and cool by emitting in the far infrared \citep[e.g.,][]{Draine2011}.  The thermal dust is 
polarized since aspherical dust grains preferentially emit along their longest axis, which tend to 
align perpendicular to magnetic fields.

In the frequency range of interest for CMB experiments, the spectrum is well approximated by a 
modified blackbody with a power-law emissivity, such that
\be
I= A \nu^{\beta_{d}}B_\nu(T_{d}),
\ee
for spectral index $\beta_d$ and temperature $T_d$, where $B_\nu$ is the Planck function. A single 
component at $T=15.9$~K fits the \planck\ data well \citep{Planck2015fgnd}, with different indices 
preferred by the intensity ($\beta=1.51\pm0.01$) and polarization ($1.59\pm0.02$) data. This 
difference indicates the presence of multiple components with different polarization properties.

In intensity the two component model of \citet{Finkbeiner1999}, with a hot and cold component at 
9.4~K and 16~K, is marginally preferred \citep{Meisner2014}. They use this model to extrapolate 
100~$\mu m$ emission and $100/240$~$\mu m $  flux ratio maps to microwave frequencies. The exact 
physical model is not well constrained by current observations, including the number of components, 
spatial variablility of spectral index, and spatial variation of the dust temperature. 

\subsubsection{Model 1: Nominal index}
\label{subsubsec:dustmodel1}

Our nominal model uses template maps at 545~GHz in intensity and 353~GHz in polarization. We use 
the templates estimated from the \planck\ data using the `{\tt Commander}' code 
\citep{Planck2015fgnd}. In polarization these maps closely match the 353~GHz \planck\ data which is 
dominated by thermal dust. We use the $N_{\text{side}}$ 2048 dust intensity map degraded to 
$N_{\text{side}}$ 512, and the polarization product smoothed to two degrees FWHM in polarization 
with small scale variations added by the procedure described in section \S\ref{sec:smallscale}.

In the nominal simulations, we model the frequency scaling as a single component, using the 
best-fit emissivity estimated by the {\tt Commander} fit. The emission model is given by
\ba
I^\text{d}_{\nu}(\vecn) &=& A_{{\rm I}, \nu_I}(\vecn)(\nu/\nu_I)^{\beta_{d}(\vecn)}B_{\nu}(T_{d}(\vecn)) \nonumber\\
\{Q^\text{d}_{\nu}(\vecn),U^\text{d}_{\nu}(\vecn)\} &=& \{A_{{\rm Q}, \nu_P}(\vecn),A_{{\rm U},\nu_P}(\vecn)\}\times  \nonumber\\
             &&  (\nu/\nu_P)^{\beta_{d}(\vecn)}B_{\nu}(T_{d}(\vecn)) 
\label{eqn:dustmodel}
\ea
Here $\nu_I=545$~GHz and $\nu_P=353$~GHz. We assume that the intensity and polarization share the 
same index, as was assumed in the {\tt Commander} fitting process. Both $\beta_{d}$ and $T_{d}$ 
vary spatially; the maps are shown in Fig \ref{fig:allmaps}.  

 This model will not capture all the of the physical complexity as it is likely that silicate and 
carbonaceous grains have distinct emissivities. They also likely have different degrees of 
polarization, since the efficiency of the grain alignment varies with the size and shape of grain. 
This would result in the polarization fraction in dust being a function of frequency, with some 
evidence for this shown in \citet{planck_dust:2015}.

\subsubsection{Models 2 and 3: Spatially variable index}
\label{subsubsec:models23}

The dust index is expected to vary spatially, in particular in polarization, but current data cannot strongly constrain this 
possible variation. We perform a test to assess how well a varying index can be detected by the \planck\ data given the 
current noise levels.

\begin{figure}
\center
\includegraphics[width=0.46\textwidth]{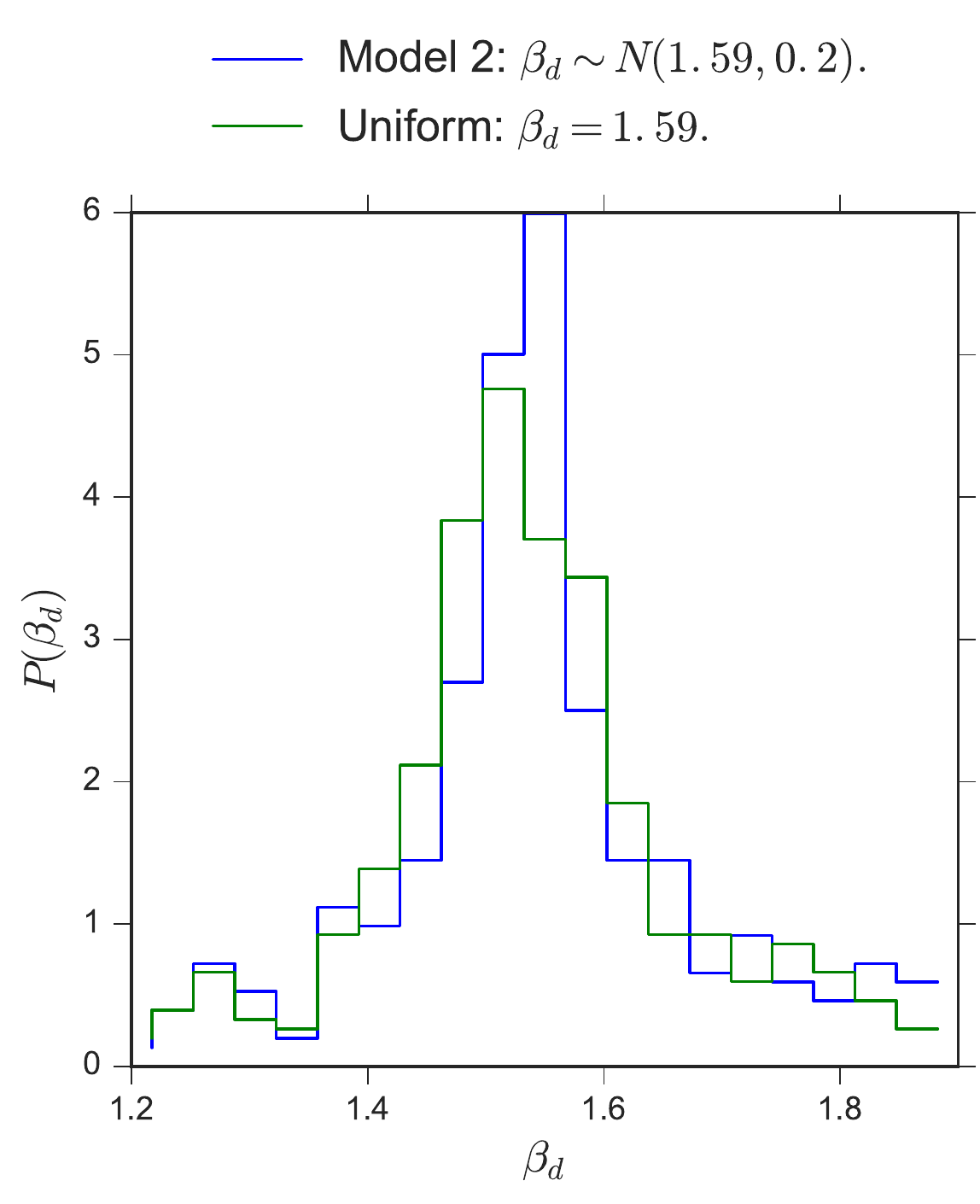}
\caption{Normalized histograms of the dust spectral index, $\beta_{d}$, calculated for noisy 
simulations of different {\tt PySM} models with varying intrinsic index dispersion.  We see that 
the resulting dispersions are very similar, indicating noise-dominated data.}
\label{fig:dust_disp}
\end{figure}

\begin{figure}
\center
\includegraphics[width=0.46\textwidth]{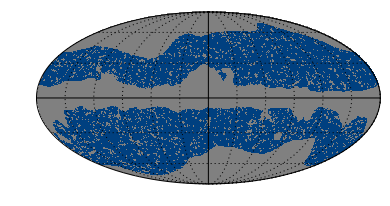}
\caption{Mask used in calculation of $\beta_{d}$ in section \ref{subsubsec:models23}. This mask is 
an approximation to the one used in the \planck\ analysis (figure 1 of \citet{planck_dust:2015}).}
\label{fig:dust_disp_mask}
\end{figure}

We simulate a spectral index map with degree-scale variation drawn from a Gaussian of mean $1.59$ and dispersion
$\sigma$. We then simulate polarized dust emission in Stokes Q and U at 217 and 353 GHz at $N_{\rm side}=128$ for 
$\sigma$ in the range 0.05 to 0.7.  We produce noise maps at 217~GHz and 353~GHz using the \planck\ half-mission 
full-sky maps at 217 and 353 GHz.  We first degrade these to $N_{\rm side}$ 128 and then at each frequency take the 
difference of the two half-mission maps and divide by a factor of 2. Finally we smooth each noise map with a 
Gaussian kernel of one degree FWHM. We then estimate the index from these maps in circles of radius ten degrees 
centred on Healpix $N_{\rm side}=8$ pixels, using

\be
\beta_{d}(\vecn) = \frac{\ln(\frac{[Q,U]_{1}(\vecn)}{[Q,U]_{2}(\vecn)}   
\frac{B(\nu_{2},T(\vecn))}{B(\nu_{1},T(\vecn))})}   {\ln(\frac{\nu_{2}} {\nu_{1}})} + 2.
\ee
This follows a similar method used in the comparable \planck\ analysis in \citet{planck_dust:2015}, 
except we do not use the 143~GHz channel and do not  add CMB and synchrotron, nor fit for them. 
We use a similar region as the \planck\ analysis, shown in Fig \ref{fig:dust_disp_mask}. The dispersion
of the indices for a uniform input index of 1.59, and for an input index map with degree-scale variation 
of standard deviation of 0.2 is shown in Figure \ref{fig:dust_disp}, and can 
be compared to Figure 9 in \cite{planck_dust:2015}. The statistics of the
recovered index distributions for these two models, the nominal model, and a model with a larger standard deviation 
of 0.3, are shown in Table \ref{table:dustdisp}.

\begin{table}
\centering{
\renewcommand*{\arraystretch}{1.4}
\begin{tabular}{@{}|p{1.7cm}|cc|@{}}
\hline
Model & Mean & Std. Dev. \\ \hline
Nominal          & 1.53 & 0.22      \\
$\sigma(\beta)=0.2$           & 1.58 & 0.24      \\
$\sigma(\beta)=0.3$         & 1.58 & 0.23      \\
Uniform          & 1.58 & 0.23      \\ \hline
\end{tabular}}
\caption{Statistics of dust polarization index calculated from different
simulations of dust polarization at 217~GHz and 353~GHz containing instrumental
noise compatible with the corresponding \planck\ channels.}
\label{table:dustdisp}
\end{table}

The distributions for $\beta_{d}$ are similar since the data are noise-dominated. The dispersion 
due to noise is $\sim0.22$ compared to the value of 0.17 found in \citet{planck_dust:2015}, and 
the value of 0.22 found in a comparable calculation by \citet{poh/dodelson:2016}. This 
indicates that models of the dust spectral index with significant spatial variation on degree 
scales are still consistent with the data.   Furthermore, a recent analysis of decorrelation of the 
\planck\ half-mission and detector set maps found an intrinsic variation of 0.07 in the dust index 
\citep{planck_dust:2016}.  Models 2 and 3 therefore modify the nominal dust model with a different 
spectral index map.  The spectral index of model 2 (3) is a Gaussian random field 
with mean of 1.59 and $\sigma = 0.2 (0.3)$ varying on degree scales for both intensity and 
polarization.

\subsubsection{Model 4: Two dust temperatures}
\label{subsubsec:dustmodel4}

One can also consider a number $N_d$ of dust components with their own temperatures and spectral 
indices:
\be\label{eq:2comp_ours}
I(\nv,\nu)=\sum_{a=1}^{N_d}\,I_a(\nv)\left(\frac{\nu}{\nu_*}\right)^{\beta_a}
\frac{B_\nu(T_a(\nv))}{B_{\nu_*}(T_a(\nv))},
\ee
and similarly for polarization. For our fourth dust model we use $N_d=2$, using the best-fit model 
templates estimated by \citet{Meisner2014} from the \planck\ data, using the model from 
\citet{Finkbeiner1999}.

The model as proposed in these references can be written as
\be
  I(\nv,\nu)=I_{\nu_0}(\nv)
  \frac{\sum_{a=1}^2 f_a\,q_a\,\left(\frac{\nu}{\nu_0}\right)^{\beta_a}B_\nu(T_a(\nv))}
       {\sum_{b=1}^2 f_b\,q_b\,B_{\nu_0}(T_b(\nv))},
\ee
where $I_{\nu_0}$ is the intensity template at $100$~$\mu$m ($\nu_0=3000\,{\rm GHz}$), $\beta_k$ 
are constant spectral indices, $T_k$ are spatially varying dust temperatures, $q_k$ is the 
IR/optical ratio for each species, $f_k$ is the fraction of power absorbed from the interstellar 
radiation field and emitted in the FIR by each component, and we have omitted the color correction 
factors.
In order to adapt this to the model in Eq. \ref{eq:2comp_ours} we generate the separate amplitude 
templates $I_a(\nv)$, at $\nu_*=545\,{\rm GHz}$ in terms of $I_{\nu_0}$ and $T_k(\nv)$ as
\be
I_a(\nv)=I_{\nu_0}(\nv)\frac{\left(\frac{\nu_*}{\nu_0}\right)^{\beta_a}\,f_a\,q_a\,B_{\nu_*}
(T_a(\nv))}{\sum_{b=1}^2f_b\,q_b\,B_{\nu_0}(T_b(\nv))}.
\ee

In polarization, we construct the polarization simulations using the polarization angles and fractional 
polarization from the 353~GHz template maps in Model 1, such that

\begin{align}
Q(\nu,\nv) &= f_d(\vecn) I(\nu,\nv) \cos(2\gamma(\nv)) \nonumber\\
U(\nu,\nv) &= f_d(\vecn) I(\nu,\nv) \sin(2\gamma(\nv)). 
\end{align}
where $f_d = \sqrt{Q^2+U^2}/I$ at 353~GHz in Model 1.

\subsection{Anomalous microwave emission}
\label{subsec:ame}

Anomalous microwave emission refers to emission with a spectral distribution not well approximated 
by known foreground models. It has been detected in compact objects, and in the diffuse sky, with early 
measurements by \citet{DeOliveira-Costa1997,Leitch1997}. It is spatially correlated with dust, and primarily 
important in the 20 - 40 GHz range, with variable peak frequency \citep{Stevenson2014}.

A likely model for the emission is rapidly spinning dust grains. \citet{Draine1998} explain the 
emission by a population of grains of size $< 3\times10^{-7}$cm, with  modest electric dipole 
moments. A candidate for these grains is polycyclic aromatic hydrocarbons (PAHs) that are detected 
in vibrational emission in the range $3 - 12 \mu m$. The theoretical SED for such spinning PAH 
grains have been successfully fit to AME observations \citep{Hoang2011}, but recent analysis of the 
\planck\ data has cast some doubt on their nature \citep{hensley/etal:2015}. A second candidate for 
AME is magnetic dipole radiation due to thermal fluctuations of magnetization in small silicate 
dust grains \citep{Draine1999}.  

\subsubsection{Model 1: Nominal unpolarized AME}

We model the AME using the \planck\ templates derived from the {\tt Commander} parametric fit to 
the \planck\ data \citep{Planck2015fgnd}, using the {\tt Commander} model:

\begin{align}
I^{\text{AME}}_{\nu}(\vecn) = &A_{\nu_{0,1}}(\vecn) \epsilon(\nu, \nu_{0,1}, 
\nu_{p,1}(\vecn),\nu_{p_{0}}) \nonumber \\ +  &A_{\nu_{0,2}}(\vecn) 
\epsilon(\nu,\nu_{0.2},\nu_{p,2},\nu_{p_{0}}).
\label{AMEmodel}
\end{align}

Here the first component has a spatially varying emissivity, and the second component a spatially 
constant emissivity. Both these emissivity functions are calculated using {\tt SpDust2} 
\citep{Ali-Haimoud2009,2011MNRAS.411.2750S}, evaluated for a cold neutral medium and shifted in 
$\log(\nu)-\log(I)$ space.  The two template maps are shown in Figure \ref{fig:allmaps}.  This 
nominal AME model is unpolarized.

\subsubsection{Model 2: Polarized AME}

AME is not thought to be strongly polarized, and the polarization fraction has been constrained to 
be below 1 - 3\% in the range 23 - 41 GHz by observations of the Perseus molecular complex 
using \wmap\ 7-year data \citep{Dickinson2011}. More recent observations of AME emission 
from the molecular complex W43 by the QUIJOTE experiment have placed a 0.39\% upper limit 
on its polarization fraction, which falls to 0.22\% when combined with \wmap\ data \citep{QUIJOTETEAM2016}.
\citet{Remazeilles2016} found that neglecting a 1\% level of polarized AME can bias the derived value 
of the tensor-to-scalar ratio by non-negligible amounts for satellite missions. 

To construct a template we use the dust polarization angles, $\gamma_{d}$, calculated from the 
\planck\ {\tt Commander} 2015 thermal dust Q and U maps at 353 GHz.  The AME polarization is then 

\be
Q_a = f_a I_{\nu}\cos(2\gamma_{353}),\hspace{12pt} U_a = f_a I_{\nu}\sin(2\gamma_{353}).
\ee
In this model we assign a global polarization fraction of 2\%; the fraction can also be easily changed 
by varying the $f_a$ parameter.

\subsection{Free-free}
\label{subsec:freefree}

Free-free emission is caused by electrons scattering off ions in the interstellar medium 
\citep{ryb}. The frequency scaling is well approximated by a function of the electron temperature 
and emission measure \citep{Draine2011}.  This is very close to a power law of $-2.14$ at 
frequencies greater than 1 GHz, and flattens abruptly at lower frequencies \citep{Planck2015fgnd}. 

Free-free has been measured in \wmap\ and \planck\ intensity data, and it should be unpolarized 
since the scattering is independent of direction.  However, there are small effects at the edges of 
dense ionized clouds due to the non-zero quadrupole moment in the electron temperature, which can 
cause up to 10\% polarization \citep[e.g.,][]{Fraisse2009}. The net polarization over the sky is 
estimated to be below 1\% \citep{Macellari2011}.

The {\tt PySM} nominal model for free-free emission assumes it is unpolarized, and uses the degree-scale
smoothed emission measure and effective electron temperature {\tt Commander} templates
\citep{Planck2015fgnd}.  We apply the analytic law presented in \citet{Draine2011} to produce an
intensity map at 30~GHz, which we then scale with a spatially constant power law index.  We choose
this index to be -2.14 consistent with \wmap\ and \planck\ 
measurements for electrons at $\sim8000$~K \citep{bennett/etal:2013, Planck2015fgnd}. This gives
\be
I^{\text{ff}}_{\nu}(\vecn) = A_{\nu_0}^{\text{ff}}(\vecn)\left(\frac{\nu}{\nu_0}\right)^{-2.14}.
\label{eq:freefree}
\ee
Different behaviour will be expected below $\sim$0.01~GHz, where the {\tt Commander} model flattens 
\citep{Planck2015fgnd}.

\subsection{CMB}
\label{subsec:cmb}

We use the {\tt Taylens} code \citep{naess/louis:2013} in {\tt PySM} to generate a lensed CMB 
realization.  The input to {\tt Taylens} is a set of $C_{l}$'s ($C_{TT}, C_{EE}, C_{BB}, C_{TE}, 
C_{\phi\phi}, C_{T\phi}, C_{E\phi}$) which have been calculated using the CAMB numerical code 
\citep{lewis/etal:2000}. The nominal model uses $\Lambda$CDM cosmological parameters that best fit the 
\planck\ 2015 data.  We incorporate the functions of Taylens into the {\tt PySM} code for 
portability, so some functionality is removed\footnote{The original code is available at 
\url{https://github.com/amaurea/taylens}}. We scale the CMB emission between 
frequencies using the blackbody function.

The user can opt to either run {\tt Taylens} during the simulation, or use a pre-computed temperature and 
polarization map supplied with the code or generated by the user. If using {\tt Taylens}, the CMB map can 
also be artificially delensed, with the expected lensing signal suppressed by a chosen factor.   

\subsection{Instrument}
\label{subsec:instrument}

We describe the instrument response with a simple top-hat bandpass, Gaussian white noise, and Gaussian
beam profile.  The user specifies a central frequency, $\nu$, and a width per band, $\Delta \nu$. The output signal is 
calculated using 
\be
I_{\nu, \Delta \nu}(   \hat{\textbf{n}}   ) = \int_{ \nu - \frac{\Delta \nu}{2}  }^{ \nu + 
\frac{\Delta \nu}{2} } \frac{ I_{\nu^{\prime}}( \hat{\textbf{n}})}{\Delta \nu} d\nu^{\prime}.
\label{eq:bandpass}
\ee
The white noise level is set per band for both intensity and polarization. The beam is characterized by a FWHM
per channel. This instrument model will not capture realistic noise realizations or realistic bandpasses; 
the code is designed to be easily modifiable to incorporate such features.

\begin{figure*}
\includegraphics[width=0.475\textwidth]{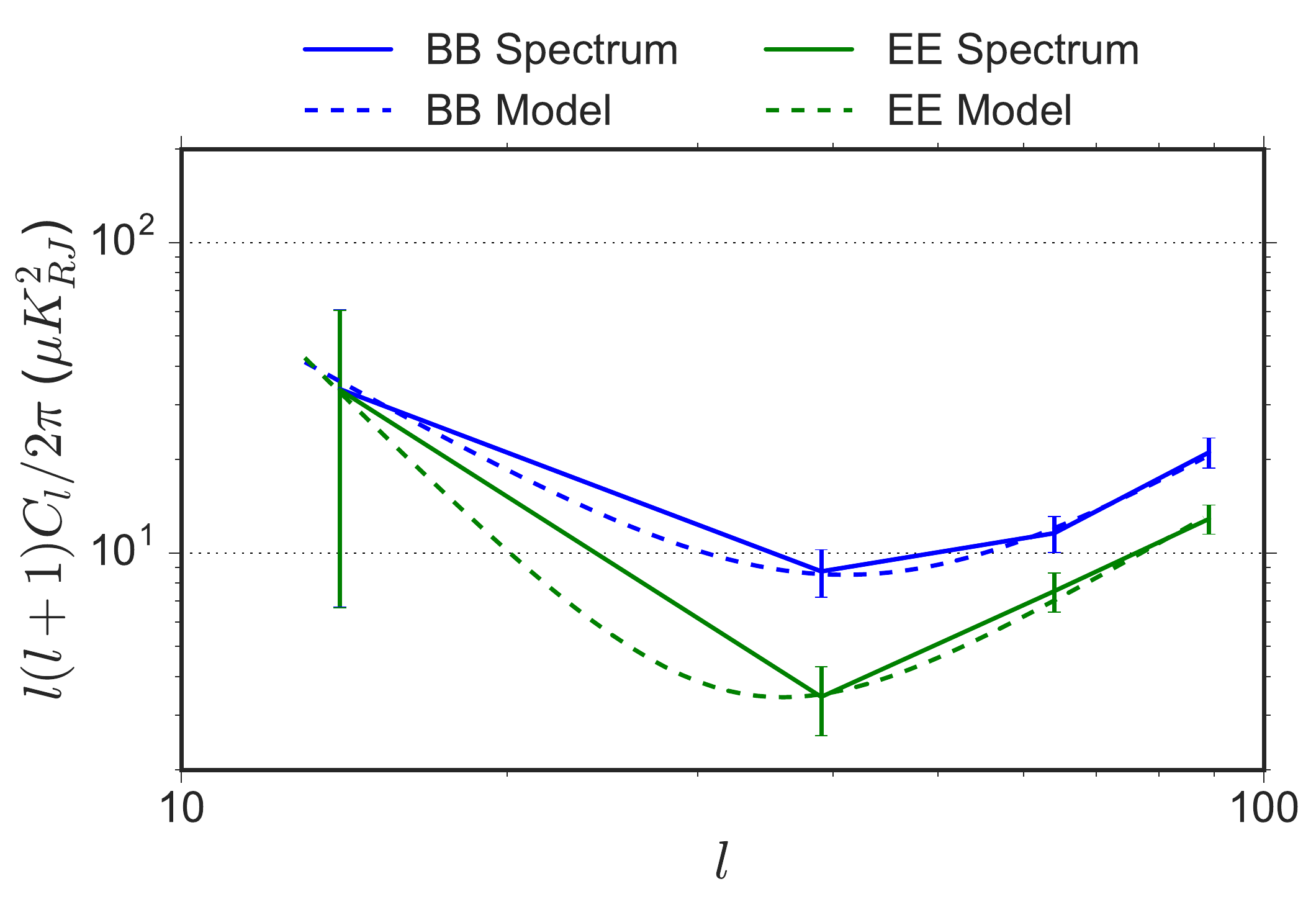}
\includegraphics[width=0.46\textwidth]{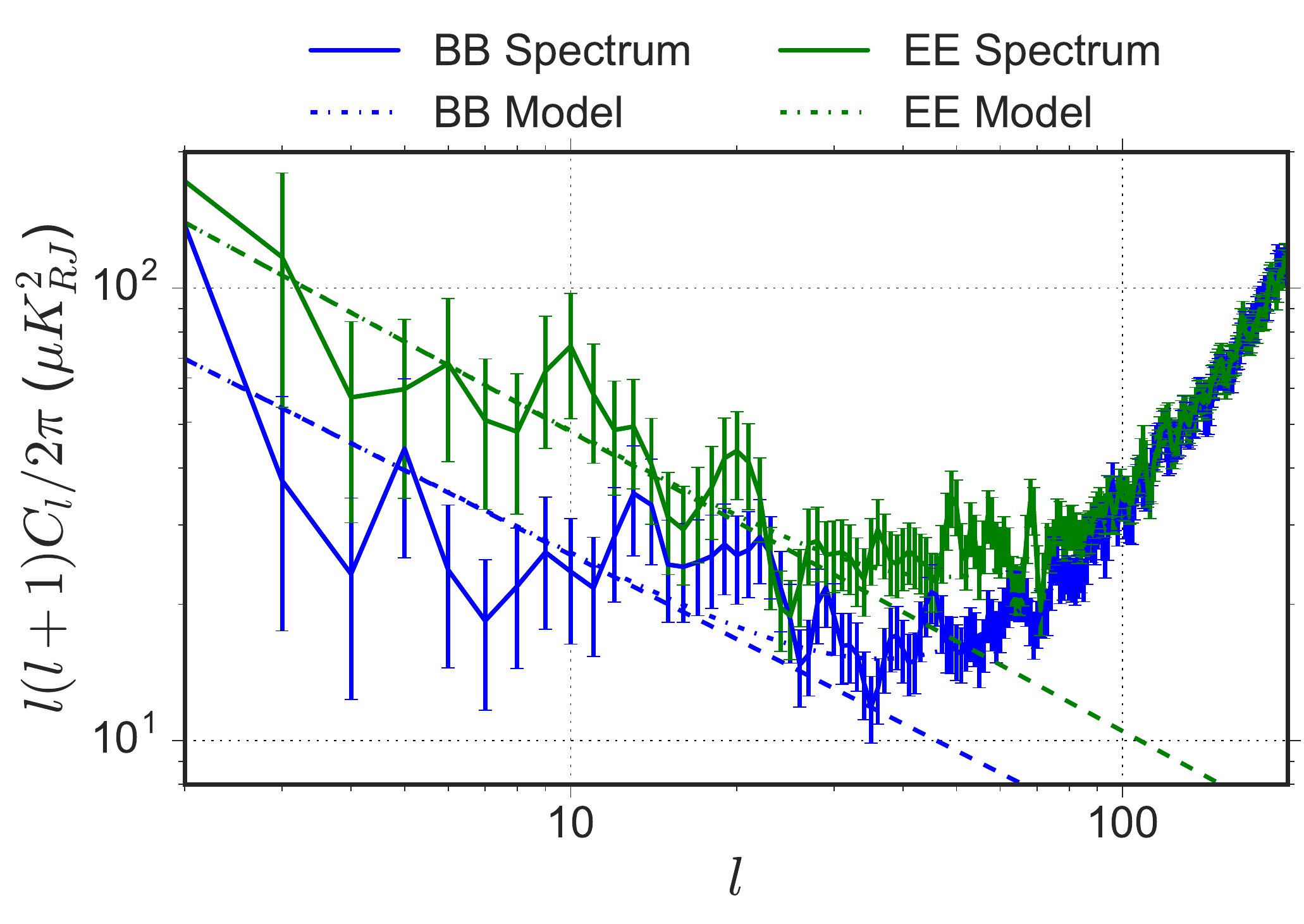}
\caption{{\sl Left}: synchrotron polarization spectra in a square region centered on 
RA, DEC = [0, -55] of size 1600 deg$^2$.  The errors shown are cosmic variance 
only.  The best-fit power-law signal plus noise model from Eqn. \ref{eq:lstarmodel} is shown.  
The BB model minimum  is used to estimate the scale $l_*$ to smooth the maps.
{\sl Right}: synchrotron polarization spectra computed with the \wmap\ polarization analysis 
mask, and best-fit model.  The dashed lines are the extrapolated power laws used in the 
small-scale simulation.}
\label{fig:synch_lstar}

\includegraphics[width=0.48\textwidth]{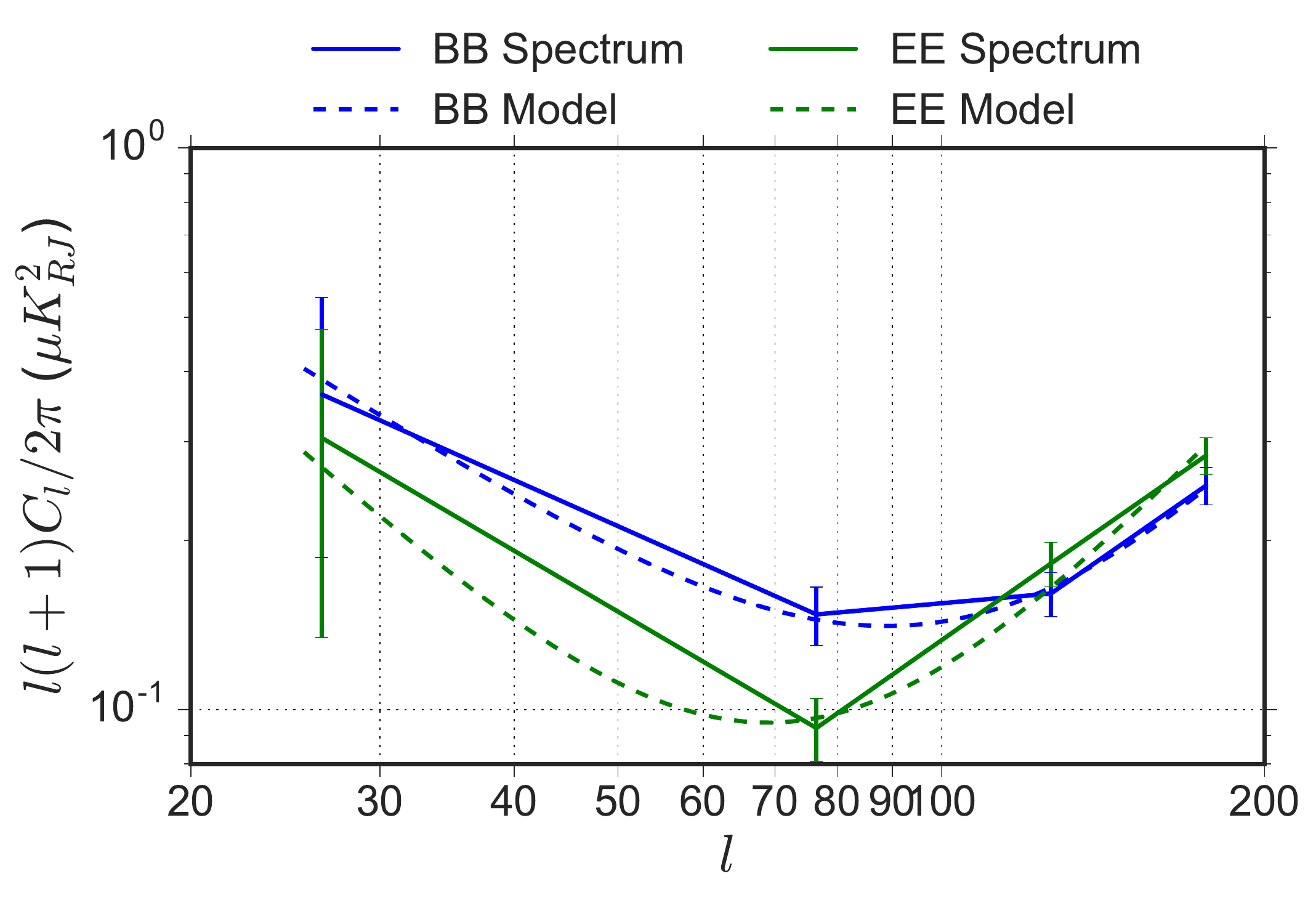}
\includegraphics[width=0.48\textwidth]{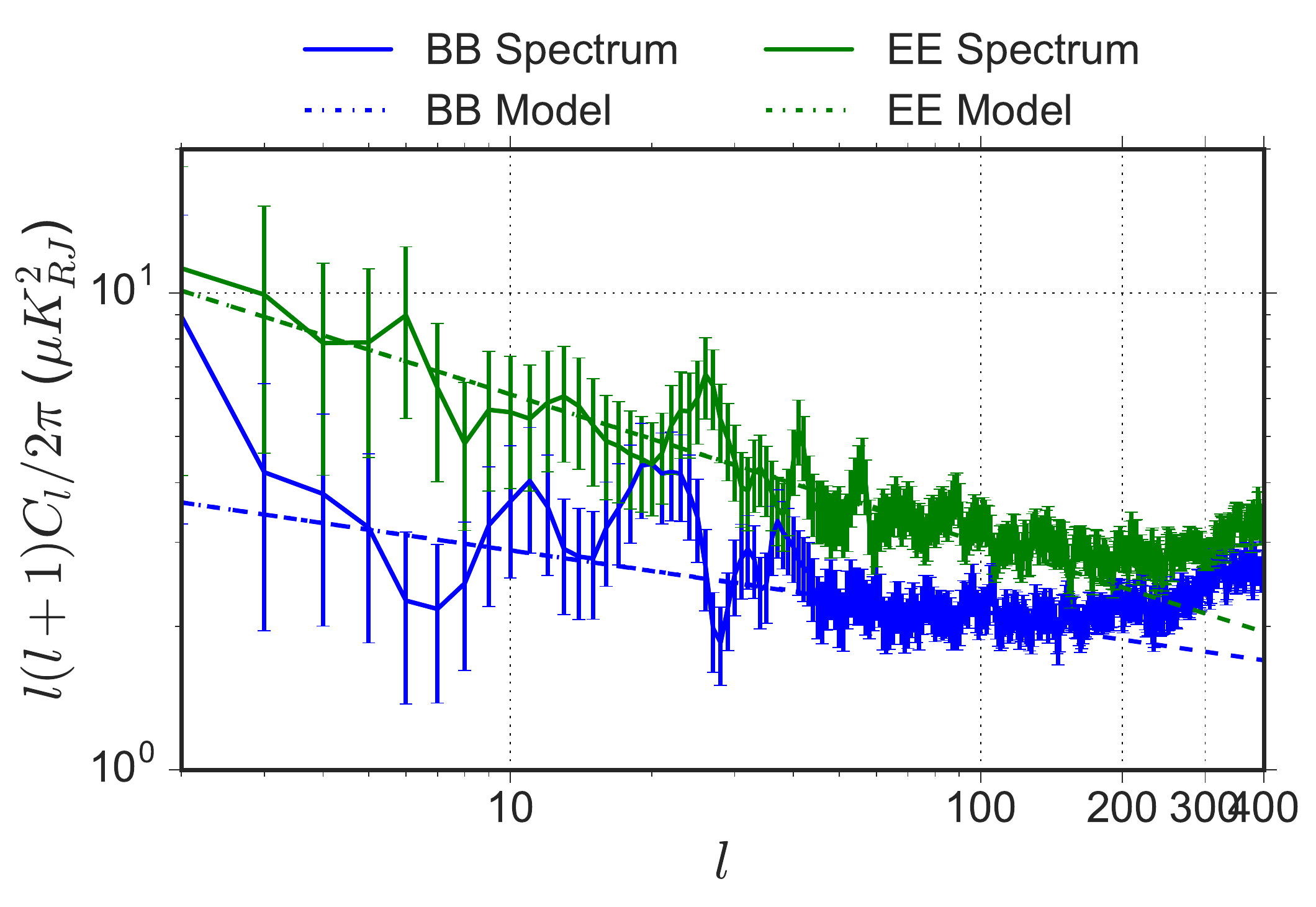}
\caption{{\sl Left}: dust polarization spectra as in Figure \ref{fig:synch_lstar}, but for a 
smaller patch of 800 square degrees.
{\sl Right}: dust polarization spectra as in Figure \ref{fig:synch_lstar}, but using the \planck\ 
Gal 80 Galactic plane mask.}
\label{fig:dust_lstar}
\end{figure*}

\begin{figure*}
\includegraphics[width=0.48\textwidth]{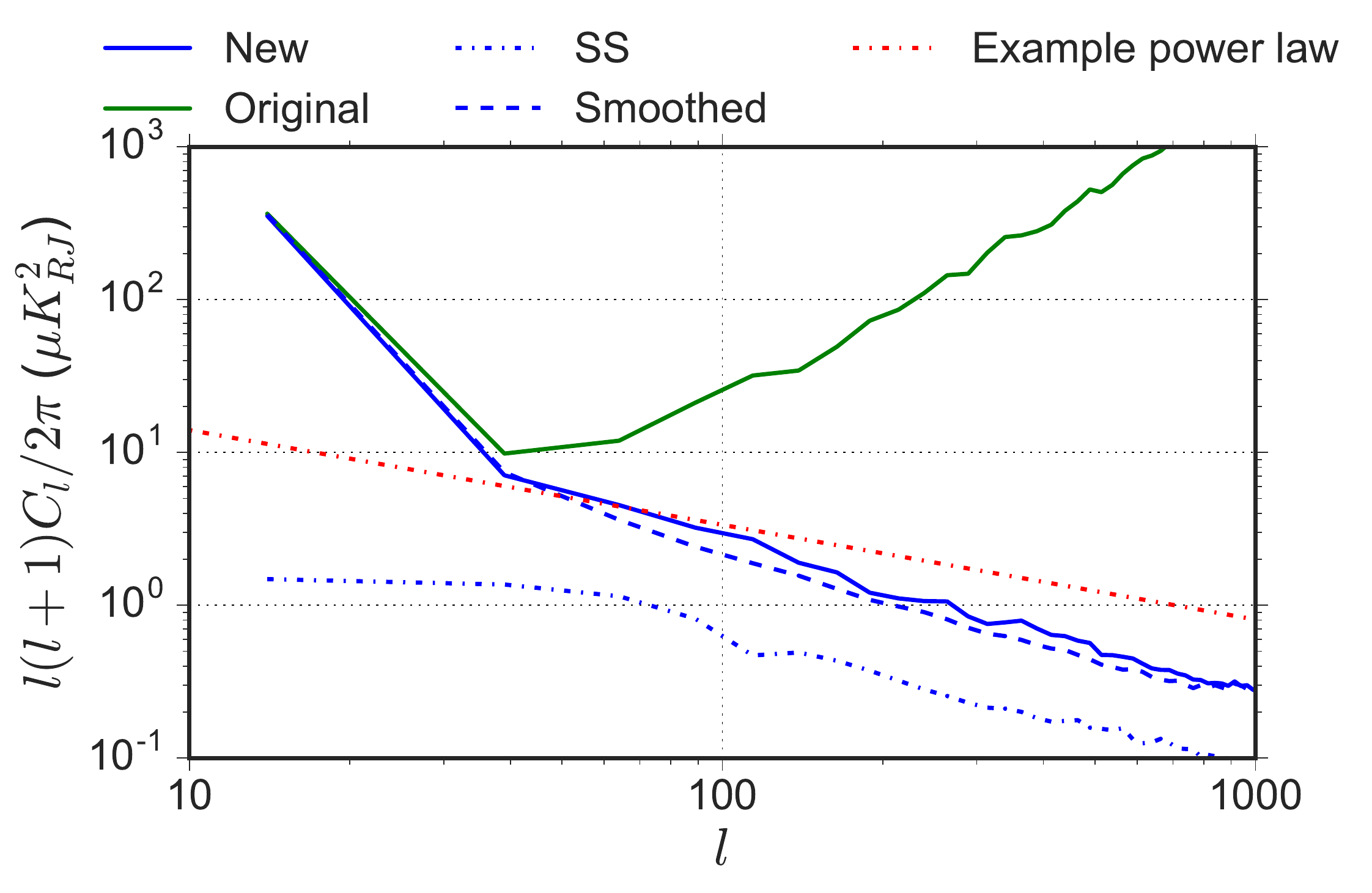}
\includegraphics[width=0.48\textwidth]{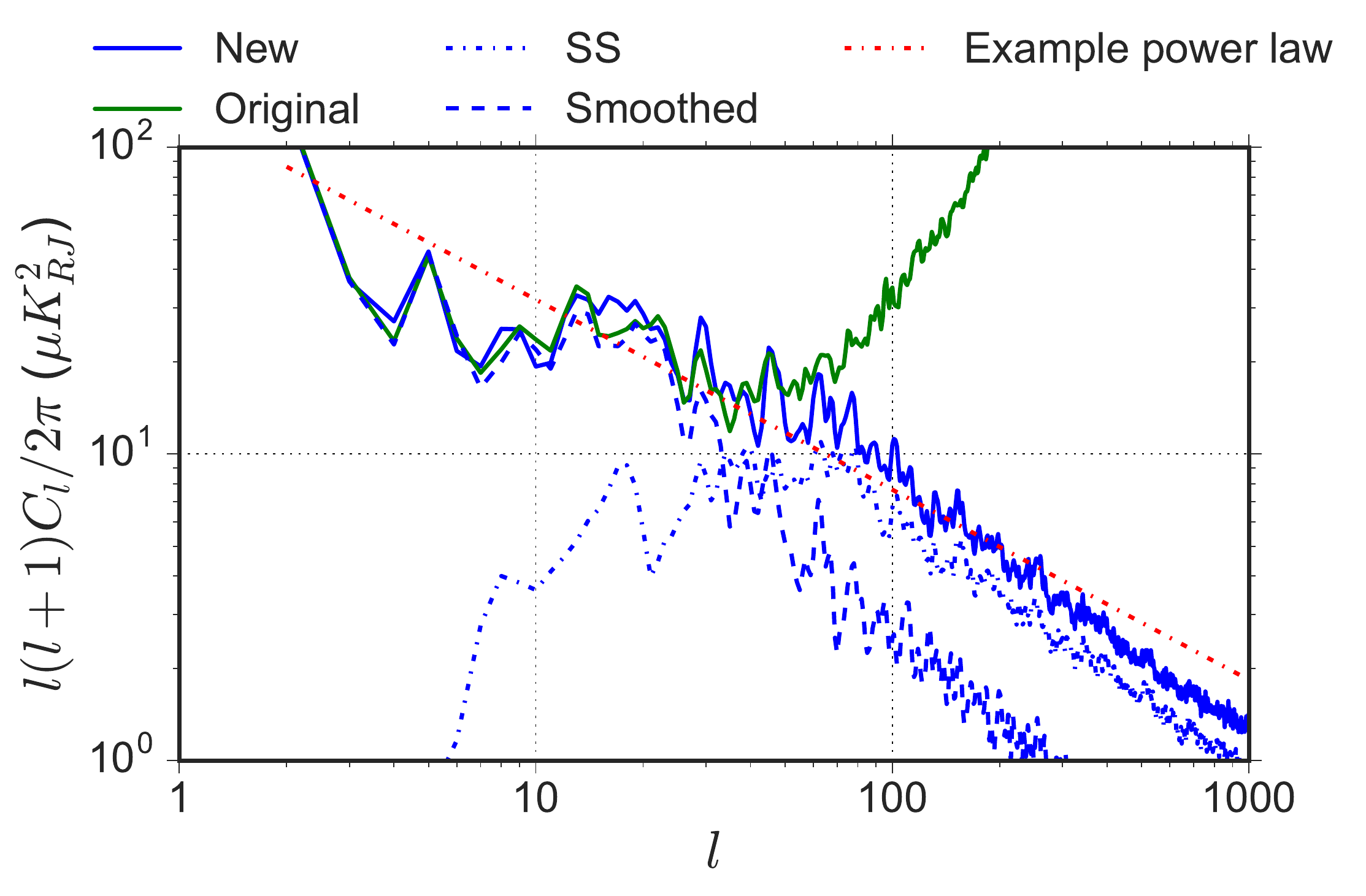}
\caption{{\sl Left}: synchrotron $BB$ spectra using the 1600 square degree region centered on RA, DEC = [0, 
-55]. We show the original template, the smoothed template, the small 
scale realization, and the final map with small scales added. The dashed red line shows the shape of the power law of the small scale realization to guide the eye.
{\sl Right}: synchrotron $BB$ spectra over 75\% of the sky using the \wmap\ polarization analysis mask. }
\label{fig:synch_final}

\includegraphics[width=0.48\textwidth]{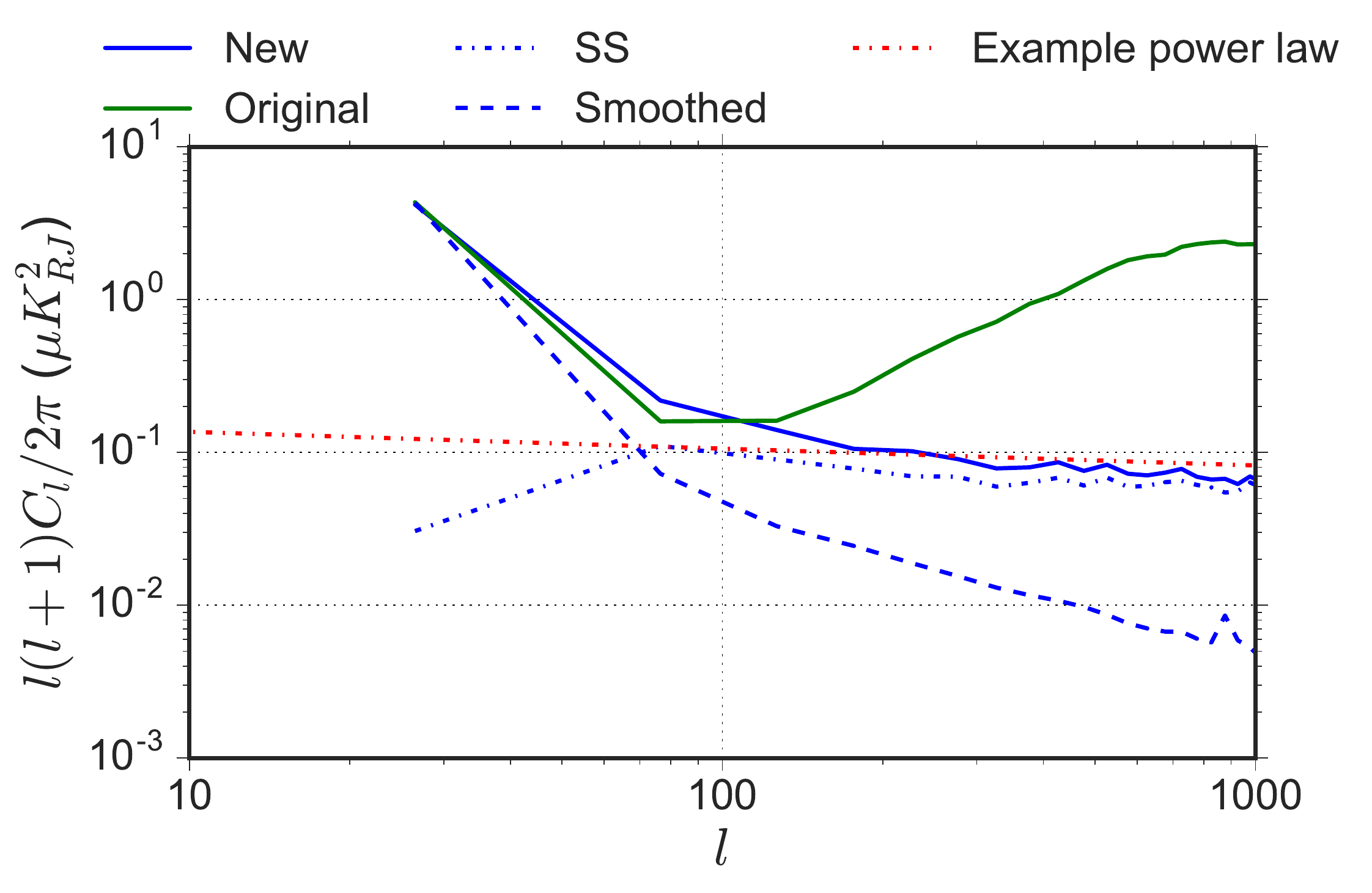}
\includegraphics[width=0.48\textwidth]{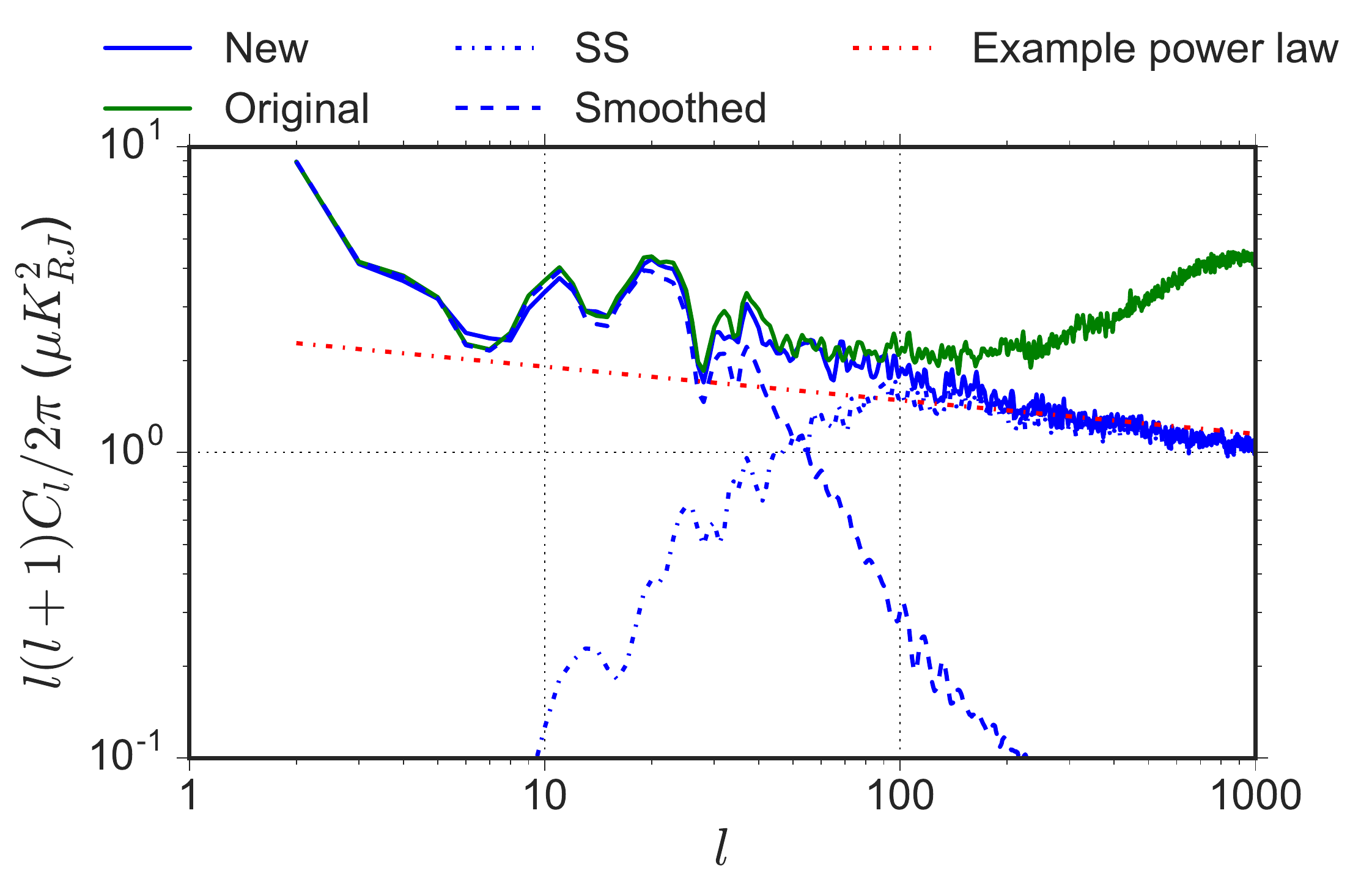}
\caption{{\sl Left}: dust BB spectra in the 825 square degree region centered on RA, DEC = [0, -55], as in Figure \ref{fig:synch_final}. 
{\sl Right}: dust $BB$ spectra in the Gal 80 region, as in  Figure \ref{fig:synch_final}.}
\label{fig:dust_final}
\end{figure*}

\begin{figure}
\includegraphics[width=0.46\textwidth]{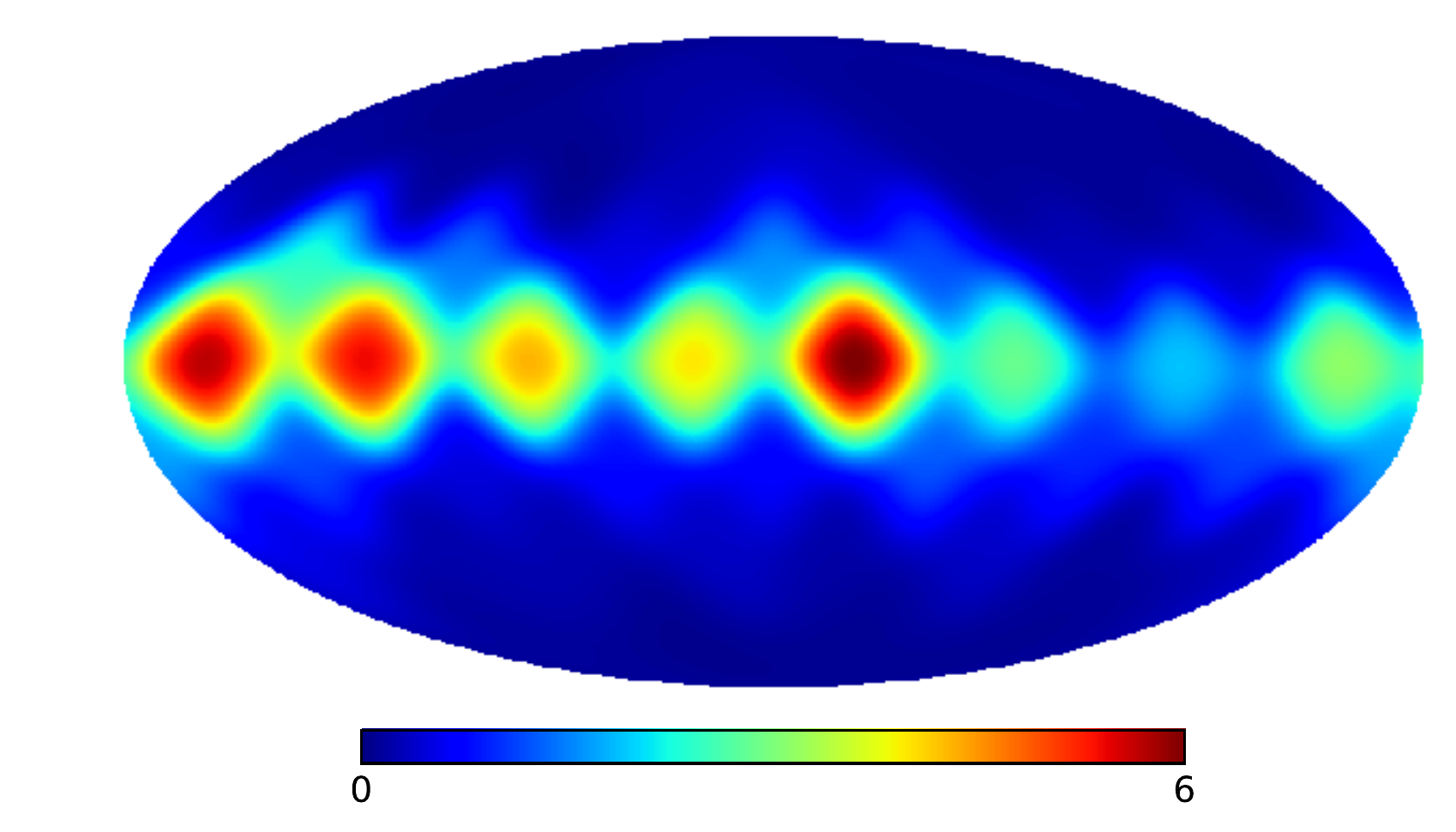}
\caption{Normalization map, $N(\vecn)$, for the dust $Q$ map.}
\label{fig:dust_norm}
\end{figure}

\section{Small-scale simulations}
\label{sec:smallscale}

Ground-based CMB experiments often observe only small patches of sky, and current data limit how well we can predict the small-scale behaviour of the foregrounds in high latitude regions at the $\ell \sim 100$ scales of interest.
Here we describe our method for simulating sky maps at a higher resolution than the available data. Our approach is to extrapolate the angular power spectrum of the available data to smaller scales, drawing a Gaussian realization from this spectrum. Other similar methods have been implemented in \citet{Miville-Deschenes2008,Delabrouille2012,Remazeilles2014a,HerviasCaimapo2016}.

We simulate intensity and polarization maps using $M = M_{0}+ M_{\rm ss}$ where $M_{0}$ is the original smoothed data and $M_{\rm ss}$ is our small-scale simulation. We implement different methods in polarization and intensity for generating $M_{\rm ss}$. Although the real sky will be non-Gaussian, we limit these small-scale simulations to Gaussian or lognormal realizations.  
   
\subsection{Polarization}   
 \label{subsec:polarizationss}

\begin{figure*}
\centering
\includegraphics[width=0.98\textwidth]{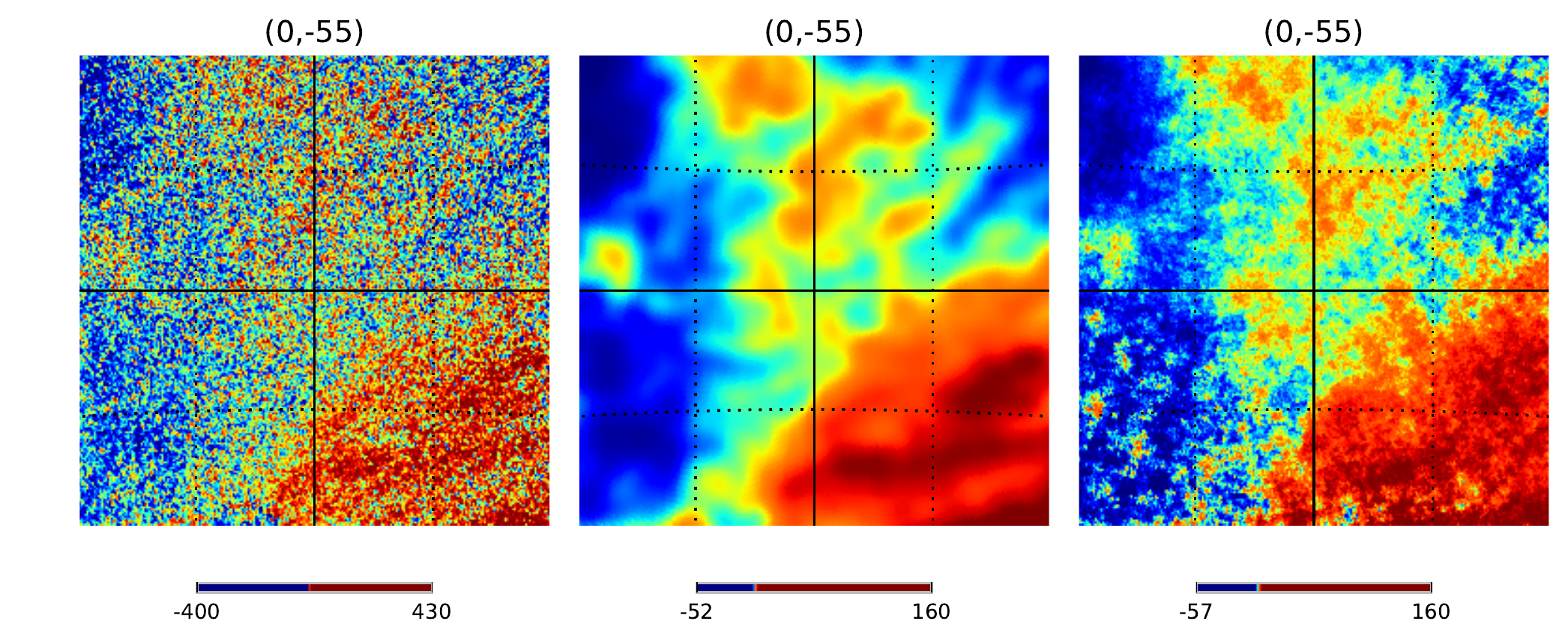}
\caption{Gnomic projection of dust $Q$ maps in a patch centered at RA, DEC = (0, -55); 40 degrees to a side. The left panel is the original map, the middle panel has been smoothed ($M_0$),
and the right hand panel has had small scales added ($M_0+M_{\text{\text{ss}}}$).  The maps are plotted in histogram-equalization in units of $\mu \text{K}_{\text{RJ}}$.}\label{fig:dust_pol_maps_ss}
\end{figure*}

 The \wmap\ and \planck\ polarization templates used in {\tt PySM} are all noise dominated at degree scales at high Galactic latitudes.  To add power to the $Q$ and $U$ maps at small-scales we determine the multipole, $\ell_*$, to which the original template is limited in resolution, smooth the maps to this scale, and add a realization of a model power law spectrum to the smoothed templates.  We compute angular power spectra on masked skies using the {\tt PolSpice} code\footnote{The {\tt PolSpice} code is available at \url{http://www2.iap.fr/users/hivon/software/PolSpice/}} \citep{Chon2004}.
 
The scale $\ell_*$ varies spatially, but here we adopt a single global $\ell_*$, which we determine by computing the polarization power spectra in a region centered on RA, DEC = [0, -55], chosen as the location of the BICEP2/Keck patch. We choose a square region of side 40 degrees for synchrotron, and 30 degrees for dust, with a larger region for synchrotron as the maps are noisier. We fit the spectra with a signal-plus-noise model,
\be
\frac{\ell(\ell+1)}{2 \pi}C^{\rm BB}_{\ell}  =  A\ell^{\gamma} + N\frac{\ell(\ell+1)}{2 \pi},
\label{eq:lstarmodel}
\ee
approximating the uncertainties on the spectrum as due only to cosmic variance. We fit for three free parameters $A$, $\gamma$ and $N$, and estimate $\ell_*$ as the scale at which this model is minimal in $BB$ or $EE$.  The masked synchrotron and dust $EE$ and $BB$ spectra are shown in Figures \ref{fig:synch_lstar} and \ref{fig:dust_lstar}. We find $\ell^{\rm{synch}}_* = 36$ and $\ell^{\rm{dust}}_* = 69$.

We generate the large-scale templates $M_0$ by smoothing the original maps with a Gaussian kernel of FWHM $\theta_{\text{fwhm}} = 180/\ell_*\,{\rm deg}$.  We then construct $M_{\rm ss}$ by assuming that the small-scales follow a power law behaviour with  $\frac{\ell(\ell+1)}{2 \pi}C^{\rm XX}_\ell  = A^{\rm XX} \ell^{\gamma^{\rm XX}}$. We find $A^{\rm XX}$ and $\gamma^{XX}$ by fitting this model to the $EE$ and $BB$ spectra calculated on the original template with a Galactic mask. We use the \wmap\ polarization analysis mask for synchrotron \citep{gold/etal:2011}, and the 80\% mask provided in the second \planck\ data, which we refer to here as Gal80. We find $\gamma^{\rm synch, EE} = -0.66, \gamma^{\rm synch, BB} = -0.62$,  $\gamma^{\rm dust, EE} = -0.31, \gamma^{\rm dust, BB} = -0.15$.

We multiply these power law spectra by the window function $1-W_\ell(\ell_*)$, where $W_\ell = \exp(-\sigma^2(\ell_*)\ell^2)$ with
$\sigma = \theta_{\rm fwhm} /\sqrt{8\ln(2)}$, such that it be added to the large-scale map that has been smoothed by the window function $W_\ell$.
We then draw a pair of $Q$ and $U$ Gaussian random fields, $\delta_G$, from this spectrum using the HEALPix\footnote{The HEALPix code is available
at \url{http://healpix.sourceforge.net}} routine \textit{synfast}.

We expect the true small-scale power to be modulated by the large-scale power, so we multiply the Gaussian random field by a spatially varying normalization such that
\be
M_{\rm SS} = N(\vecn)\delta_{G}(\vecn).
\ee

We choose $N(\vecn)$ by dividing the sky into Healpix $N_{\rm side} = 2$ pixels and computing the  angular power spectrum in each patch, $C_{\ell}(\vecn)$, and smoothing this map with FWHM $10^{\circ}$ to avoid sharp pixel boundaries.  We define
\be
N(\vecn) = \sqrt{ \frac{ C_{l_*} (\vecn) } {A \ell_*^\gamma}} 
\ee
so that the small-scale realization is normalized by the large-scale power in each patch. The $N(\vecn)$ for the dust $Q$ template is shown in Figure \ref{fig:dust_norm}.

A patch of the resulting $Q$ map for dust is shown in Figure \ref{fig:dust_pol_maps_ss}, illustrating the large-scale and additional small scale components. We also show the power spectra of the maps in Figures \ref{fig:synch_final} and \ref{fig:dust_final}, both for the masked all-sky maps and the smaller regions centered at [0,-55]. In both regions the power law behaviour is continuous at $\ell = \ell_*$.

We note that a limitation of this method is that it does not capture spatial variations in the modulation of the small-scale signal on scales smaller than $N_{\rm side}$ 2 pixels, so the normalization will not be accurate in these small regions.

\subsection{Intensity: Synchrotron}
\label{subsec:lognormal}

We use a similar procedure for simulating the intensity at small-scales, but we use a lognormal rather than Gaussian distribution because it guarantees that the final map will be positive.  It is also possible to generate a lognormal distribution from a Gaussian random field, and maintain the shape of the Gaussian field's angular power spectrum to a good approximation. In these simulations we do not impose a correlation between the intensity and polarization at small scales.

For synchrotron, the Haslam template is provided at 57 arcminute resolution, which defines $M_0$. As for the polarization we fit a power law to the signal, finding $\gamma= -0.55$. We draw a Gaussian realization $\delta_G$ with variance $\sigma_G^2$, but here we generate $M_{\rm{ss}}$ using a lognormal distribution with

\be
M_{\rm ss} = M^{\rm min}_0 [ \exp(R( \vecn ) \delta_{\rm G}(\vecn)-\sigma^2_{\rm G}/2)-1].
\label{eq:lognormal}
\ee

Here $R(\vecn)$ normalizes the small scales. Instead of using the local power spectrum, we normalize the small-scale intensity map by the large-scale intensity smoothed to 4$^\circ$ and raised to a power

\begin{align*}
R(\vecn) &= \left[\frac{M_{0}(\vecn) }{\langle M_{0} \rangle}\right]^{\alpha}.
\end{align*}

\begin{figure}
\includegraphics[width=0.48\textwidth]{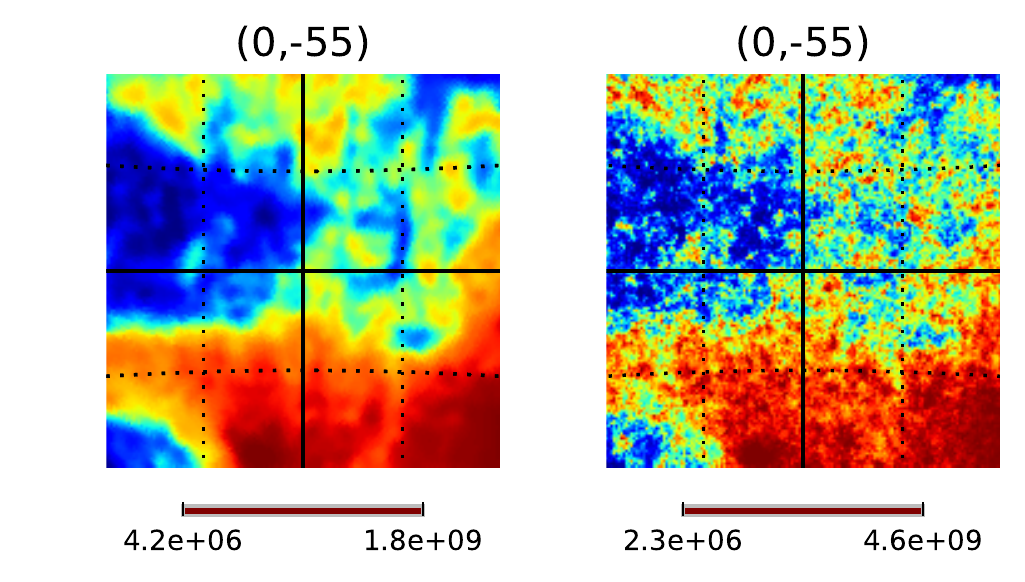}
\includegraphics[width=0.48\textwidth]{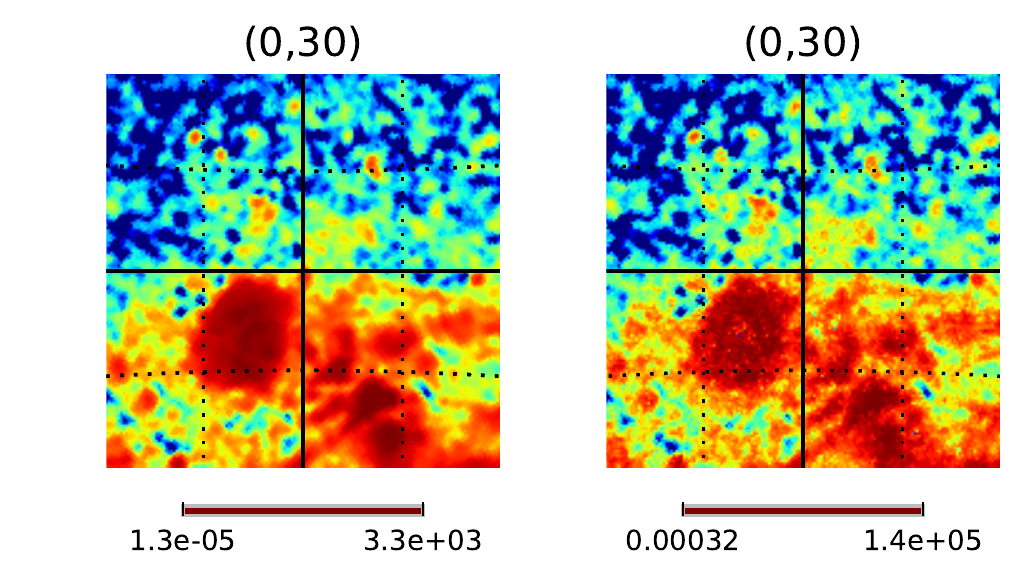}
\includegraphics[width=0.48\textwidth]{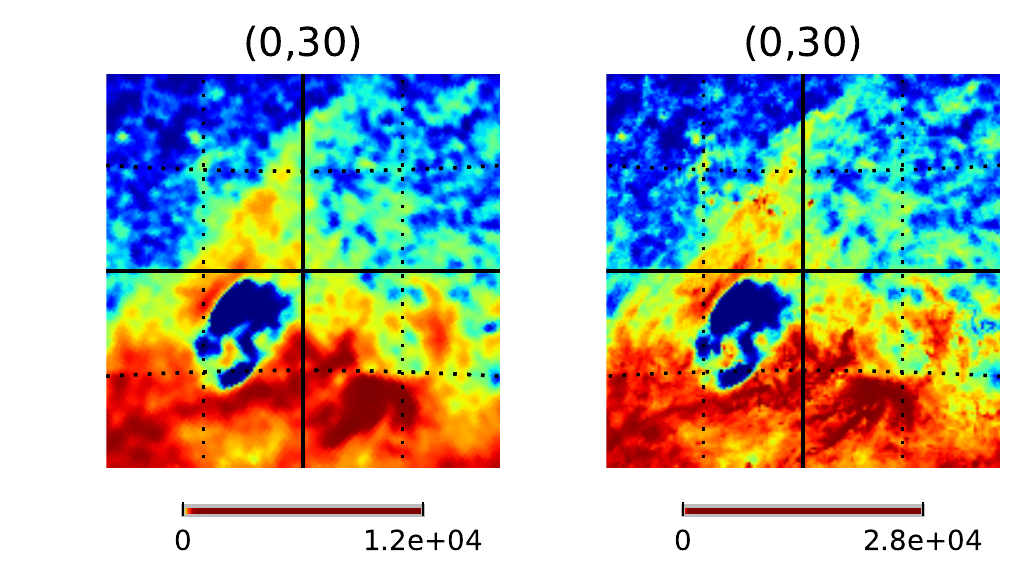}
\caption{Synchrotron (top), free-free (middle), and AME (bottom) simulated intensity maps in a patch of side 40$^\circ$ centered at RA, DEC as indicated. Left: original template; right: simulation including small scales. These have been plotted in histogram-equalization to increase the dynamic range.}
\label{fig:patch_map}
\end{figure}

We find the best-fit $\alpha=0.6$ that results in a total power spectrum of $\ell(\ell+1)C_\ell \propto \ell^\gamma$, fit in the multipole range $200<\ell<1000$.
An example of the synchrotron maps are shown in Figure \ref{fig:patch_map}.

\subsection{Intensity: Free-free}
\label{subsec:ffss}

The free-free template is smoothed at degree scales, which defines $M_0$. We found the lognormal procedure to be unsuitable for generating small scales for the free-free maps, as the comparatively larger dynamic range in small patches caused the exponential term to yield unrealistically large variation on small scales.
We also found the free-free angular spectrum to be flatter than the synchrotron, so a direct extrapolation of the power law to smaller scales produced excess power at small scales that is likely not physical.
We therefore fixed the gradient of the free-free power spectrum to be $\gamma=-0.5$, and used this to generate a $\delta_G$ realization with variance $\sigma_G^2$.  We then take the small scale map to be: 
\be
M_{\rm{ss}}(\vecn) =  R(\vecn)\delta_{\text{G}}(\vecn)
\ee
where $R(\vecn) = \langle M_0 \rangle(M_0(\vecn)/\langle M_0 \rangle)^\alpha/4 \sigma_{\text{G}}$.
We find that $\alpha = 1.15$ is the best-fit value to recover the correct power law behaviour of the power spectrum in the range $200<\ell<1000$.  We redrew $\delta_{\text{G}}$ for any negative pixels from additional full-sky realizations until we have positive values everywhere. This was necessary for $< 0.5 \%$ of pixels. An example is shown in Figure \ref{fig:patch_map}.

\subsection{Intensity: Thermal dust and anomalous microwave emission}
\label{subsec:ametdss}

The \planck\ thermal dust map has a power spectrum in the low-foreground [RA, DEC = 0, -55] region that falls off approximately as a power law. This indicates that the thermal dust intensity map is signal dominated in this high Galactic latitude region at small scales, and we do not add additional components.

The AME templates are limited to degree resolution, so we use the high resolution thermal dust product as a proxy for the AME small scales.  We produce the final AME map by multiplying the two intensity templates by the ratio of the high resolution thermal dust template and the dust template smoothed to one degree FWHM.  An example is shown in Figure \ref{fig:patch_map}. The resulting AME templates therefore have the same small-scale morphology as the thermal dust template.  Since the AME polarization templates are produced from the thermal dust polarization products we do not simulate AME polarization separately.  

\section{Discussion}
\label{sec:discussion}

We have presented new software to simulate the Galactic microwave sky in polarization and intensity.  The nominal models reflect the current understanding of Galactic foregrounds, and we have included a set of simple alternative models that capture physical extensions to these models and are still consistent with current data. There are many more possible alternatives that are not included, but we provide the public code in a way that makes adding further astrophysical complications straightforward.  The code is also fast, portable, and easy to install and begin using.

We have developed methods for the addition of simulated small scale variation in polarization and intensity, recovering power law behaviour of the polarized components in sky patches of low and high signal, with minimal noise biasing. These simulations may aid in forecasting for ground-based observations limited to partial sky coverage.  
These small-scale simulations have certain limitations. Different simulated components are not correlated, and the small-scale procedure loses information in high signal-to-noise regions by smoothing at a single scale. Incorporating the spatially varying signal-to-noise into the definition of this smoothing scale would provide more accurate simulations. The small-scales will also be non-Gaussian in practice, which we do not account for.  

There are other approaches to foreground modelling. {\tt PySM} uses 2D sky maps and parametric models to extrapolate single frequency maps to different frequencies.  This will be limited in its ability to replicate the polarized nature of Galactic foregrounds.  Due to the combination of the complex three-dimensional structure of the Galaxy's magnetic field and the stacking of different sources along any given line of sight we may expect the polarization fraction of any component to be a function of frequency.  Even on a microphysical level there is good evidence that the polarization spectrum of thermal dust is frequency dependent \citep[e.g.][]{planck_dust:2015}, as carbonaceous and silicate grains may align with the Galactic magnetic field with different efficiencies. More realistic simulations could be derived from three dimensional realizations of the Galaxy's magnetic field and source distributions.  

\section*{Acknowledgments}

BT acknowledges the support of an STFC studentship; JD and DA acknowledge the support of ERC grant 259505. DA acknowledges support from BIPAC. We thank Sigurd N\ae ss for useful comments and for use of the {\tt Taylens} code within {\tt PySM}.  We acknowledge use of the \wmap\ public maps on LAMBDA, the \planck\ public maps on the Planck Legacy Archive, the HEALPix software and analysis package \citep{gorski/etal:2005}, and the Planck Sky Model code \citep{Delabrouille2012}. 

\FloatBarrier

\setlength{\bibhang}{2.0em}
\setlength\labelwidth{0.0em}
\bibliography{library}
\end{document}